\documentclass[11pt]{article}

\newtheorem{lemma}{Lemma}
\newtheorem{proposition}{Proposition}
\newtheorem{theorem}{Theorem}

\begin{document}
\title{Local thermal equilibrium and\\
ideal gas Stephani universes}

\author{Bartolom\'{e} C{\sc oll}$^1$ and Joan Josep F{\sc errando$^2$}}
\date{\today}

\maketitle

\vspace{2cm}
\begin{abstract}
The Stephani universes that can be interpreted as an ideal gas
evolving in local thermal equilibrium are determined. Five classes
of thermodynamic schemes are admissible, which give rise to five
classes of regular models and three classes of singular models. No
Stephani universes exist representing an exact solution to a {\em
classical} ideal gas (one for which the internal energy is
proportional to the temperature). But some Stephani universes may
approximate a classical ideal gas at first order in the
temperature: all of them are obtained. Finally, some features
about the physical behavior of the models are pointed out.

\end{abstract}
\vspace{0.5cm} KEY WORDS: thermodynamic, inhomogeneous
cosmological models
\vspace{2cm}\\
$^1$ Syst\`emes de r\'ef\'erence relativistes, SYRTE-CNRS,\\
Obsevatoire de Paris, 75014 Paris, France.\\
E-mail: \texttt{bartolome.coll@obspm.fr}\\
\vspace{0cm}\\
$^2$ Departament d'Astronomia i Astrof\'{\i}sica, \\Universitat
de Val\`encia, 46100 Burjassot, Val\`encia, Spain.\\
E-mail: {\tt joan.ferrando@uv.es}

\newpage

\section{\large INTRODUCTION}

The second member of Einstein equations describes the energy
content of the medium, but not the medium itself. So, exact
solutions to Einstein equations correspond to prescribed forms of
energy tensors, but they do not fix the nature of the particular
(class of) physical fluid(s) that creates them.

Generically \cite{generically}, the physical interpretation of an
exact solution to Einstein equations, needs more than that of
their {\em energy quantities}, i.e. more than that of the
quantities that appear in the energy tensor: it is also necessary
to 'close the context'. Let us see what this means.

A non vacuum physical space-time is the history of {\em one} of
the possible evolutions of a physical medium. But there are in
general many physical media that can have in common one or some
particular evolutions (for example, many fluids admit a rigid and
static evolution with same energy density an pressure). In fact,
from the evolution point of view, the definition of a medium is
tantamount to the knowledge of the set of {\em all\,} its possible
physical evolutions. Of course, this knowledge does not need to be
{\em extensive}, it is sufficient that it be {\em intensive,} i.e.
that it be done by the {\em set of equations} whose space of
solutions is precisely the whole set of all its possible physical
evolutions.

    But according to a universal determinism requirement in
Physics, such a set of equations has to be (causally) {\em
closed}, i.e. has to admit a well posed initial value problem.

 In general the conservation equations imposed by Einstein
equations on the energy tensor (energy conservation equations) do
not constitute a closed system of equations for the energy
quantities. This is why we say that their physical interpretation
is not sufficient to define the medium. To 'close the context'
means here just to choice a {\em closure} for the energy
conservation equations, i.e. to complete these last ones with some
other physically meaningful equations in such a way to obtain a
closed system.

The closure of the energy conservation equations usually
\cite{thermoclosures} obliges to introduce new quantities, related
at least in part to the Duhem-Gibbs balance equation, and called
by this fact {\em thermodynamic quantities}.

Thus, for a {\em perfect energy tensor},
\begin{equation}
\label{energytensor} {\rm T} = (\rho + p) u \otimes u - p\,g \, ,
\end{equation}
where $\rho$ is the {\em energy density}, $p$ the {\em pressure}
and $u$ the (unit) {\em velocity} of the fluid, the usual closure
\cite{usual closure} corresponds to the hypothesis of {\em local
thermal equilibrium} (also called {\em standard thermodynamic
scheme}). The local thermal equilibrium evolution of a fluid is
thus determined by the following four conditions:
\begin{itemize}
\item Energy-momentum conservation: $\ \ \nabla \cdot {\rm T} = 0$.
\item The energy density $\rho$ is decomposed in terms of the
{\em matter density} $r$ and the {\em specific internal energy}
$\epsilon$: $\ \ \rho= r(1+\epsilon)$.
\item The equation of conservation of matter is required:
$\ \ \nabla \cdot (ru) = 0$.
\item The thermodynamic quantities {\em temperature} $T$
and {\em specific entropy} $s$ are related by equations of state
compatible with the thermodynamic Duhem-Gibbs relation: $\ \ T ds
=  d\epsilon + p d (1/r) $.
\end{itemize}

It is clear that this standard thermodynamic closure is a strict
constraint on the whole space of solutions to the Einstein
Equations for perfect energy tensors, i.e. {\em not all} the
solutions of this last space can be interpreted as being local
thermal equilibrium evolutions of a perfect fluid  \cite{perfect}.

Observe that this constraint notably increases the number of
variables because it involves not only the starting energy
quantities \,($u,$ $\rho,$ $p$)\, but also the new thermodynamic
ones \,($r$, $\epsilon,$ $T,$ $s$). So that, an important question
arises: {\em is it possible to know, by the sole inspection of the
energy quantities of a perfect energy tensor, if it describes the
local thermal equilibrium evolution of a perfect fluid?}

As we showed some years ago \cite{cf2}, the answer is positive.
Its precise formulation (denoting by $u(x)$ the directional
derivative, with respect to $u$, of the quantity $x$) is given by
the following

\vspace{2mm}
\noindent
{\bf Theorem $\chi$ (local thermal
equilibrium):} {\em A divergence-free perfect energy tensor
evolves in local thermal equilibrium if, and only if, the
space-time function $\chi \equiv u(p)/u(\rho)$ depends only on the
variables $p$ and $\rho$ : $d\chi \wedge dp \wedge d\rho = 0.$}

\vspace{2mm}

This result is strikingly interesting because it shows that, in
spite of its name, of its historic origin and of its usual
conceptualization, the notion of {\em local thermal equilibrium}
for a perfect fluid is a {\em purely hydrodynamic}, not
thermodynamic, notion \cite{jerarquia}.

The function $\chi$ is called the {\em  indicatrix of local
thermal equilibrium}. To every physical fluid in local thermal
equilibrium, whatever its evolution be, corresponds a sole,
particular, indicatrix (particular equation of state) which
physically represents the square of the sound velocity in this
fluid \cite{cf3}. So that the direct application of theorem $\chi$
not only allows

 \indent \indent {\em i)} to detect, among the known families of
solutions to Einstein equations for a perfect energy tensor, those
sub-families that describe evolutions in local thermal equilibrium
of physical fluids, but also

 \indent \indent {\em ii)} to separate these last sub-families in
sub-sub-families that describe different (local thermal
equilibrium) evolutions of {\em the same} physical fluid.

Conversely, theorem $\chi$ generates the first general method to
know wether a solution for a perfect energy tensor is the
evolution of a previously given fluid: it is sufficient to verify
if the indicatrix function $\chi$ of the solution coincides with
the square of the thermodynamic expression of the sound velocity
of the given fluid expressed as a function of the variables $p$
and $\rho$. In \cite{cf3} we pointed out these applications and
gave a concrete example studying the thermodynamic class II
Szekeres-Szafron space-times.

In the present paper we consider, on one hand, the paradigmatic
{\em ideal gas} as the previously given particular fluid and, on
the other hand,  the Stephani universes \cite{st} as starting
family of solutions to Einstein equations for a perfect energy
tensor, both because of their  relative generality and their
frequent applications in cosmology and stellar interiors.

The Stephani universes admitting a generic thermodynamic scheme
were obtained by Bona and Coll \cite{bc} and later revisited by
Krasi\'nski et al. \cite{ks}. In a recent paper, Sussman \cite{su}
has analyzed a family of spherically symmetric Stephani universes
that may be interpreted as either a classical mono-atomic ideal
gas or a matter-radiation mixture.

Here we are interested in the deductive and systematic
determination of {\it all}\, Stephani universes that represent the
evolution in local thermal equilibrium of a {\em generic} ideal
gas and, among them, of those that admit a first order contact in
the temperature with the {\em classical} ideal gas
\cite{defidealgas}.

The main results of this work are based on previous ones: firstly,
on our hydrodynamic characterization of an ideal gas in local
thermal equilibrium \cite{cf}, that we shorten here in section 2;
secondly, on the above cited work by Bona and Coll \cite{bc} about
the thermodynamic schemes in Stephani universes, that we summarize
in section 3. In section 4 we apply an algorithm introduced in
section 2 to a canonical form for the thermodynamic Stephani
universes presented in section 3 in order to obtain: i) a theorem
(theorem 1) that characterizes the generic ideal gas Stephani
universes and ii) the associated indicatrix function and five
classes of compatible ideal gas thermodynamic schemes. In section
5 we integrate, up to a Friedmann-like equation, the ideal gas
equations obtained in theorem 1; this integration leads to
different models with thermodynamics of a specific class. In
section 6 we study the Stephani ideal gas models that approximate
a classical ideal gas at first order in the temperature. Finally,
section 5 is devoted to analyze how complementary physical
requirements, like energy and compressibility conditions, restrict
the domain where the models have a good physical behavior. A part
of the results of this paper were stated without proof in
\cite{cf4}.

\section{\large LOCAL THERMAL EQUILIBRIUM EVOLUTION OF AN IDEAL GAS}

As already mentioned, Theorem $\chi$  provides an easy and general
method (in fact the first known one) to answer if whether or not a
solution to Einstein equations for a perfect energy tensor
corresponds with a given thermodynamic fluid: to check that, as
functions of $p$ and $\rho,$ the indicatrix function of the
solution and the square of the sound velocity of the fluid
coincide.

Elsewhere \cite{cf} we have given the hydrodynamic
characterization of the sound velocity for some representative
classes of fluids. We present here in two propositions the part of
these results about ideal gases that we will use in this paper.

Remember that a perfect energy tensor is said {\em barotropic} if
$p$ and $\rho$ are not independent, $dp \wedge d\rho = 0$, and
that it is said {\em isoenergetic} if $u(\rho) = 0.$ As we shall
see in the next section, for the class of space-times analyzed
here, the Stephani universes, the barotropic case leads to the
already well known Friedmann-Robertson-Walker models, meanwhile
the isoentropic case is not possible because of the non vanishing
expansion of these fluids.

A (generic) ideal gas satisfies the equation of state $p=krT$. For
it we have shown the following hydrodynamic characterization
\cite{cf}: \vspace{0.1cm}
\begin{proposition}  \label{pro:idealgas1}
The necessary and sufficient condition for a non barotropic and
non isoenergetic divergence-free perfect energy tensor {\rm
(\ref{energytensor})} to represent the local thermal equilibrium
evolution of an {\rm ideal gas} is that its indicatrix function
$\chi\equiv u(p)/u(\rho)$ be a non identity function of the
variable $\pi \equiv p/\rho$:
\begin{equation}
\label{dchi-wedge-dpi}
d \chi \wedge d\pi = 0 , \qquad \chi \not= \pi   \label{xi-pi}
\end{equation}
\end{proposition}

\begin{proposition} \label{pro:idealgas2}
A non barotropic and non isoenergetic divergence-free perfect
fluid energy tensor verifying {\em (\ref{xi-pi})} represents the
local thermal equilibrium evolution of the ideal gas with specific
internal energy $\epsilon$, temperature $T$, matter density $r$,
and specific entropy $s$ given by
\begin{eqnarray}
\epsilon(\rho,p) = \epsilon(\pi) \equiv e(\pi)-1 \, , \qquad \quad
T(\rho,p) = T(\pi) \equiv {\pi \over k} e(\pi) \, , \label{T} \\
r(\rho,p) = {\rho \over e(\pi)} \, ,  \qquad  \qquad    \qquad  \
\quad s(\rho,p) = k \ln \frac{f(\pi)}{\rho} \, , \quad  \quad \ \
\label{r-s}
\end{eqnarray}
$e(\pi)$ and $f(\pi)$ being, respectively,
\begin{equation}
e(\pi) = e_0 \exp\{\int \psi(\pi)d\pi \} \, , \qquad  \qquad \psi
(\pi) \equiv \frac{\pi}{(\chi(\pi)-\pi)(\pi+1)} \, ; \label{e-pi}
\end{equation}
\begin{equation} \label{f-pi}
f(\pi) = f_0 \exp\{\int \phi(\pi)d\pi\} \, , \qquad \qquad
\phi(\pi) \equiv {1 \over \chi(\pi)-\pi} \, . \qquad \qquad
\end{equation}
\end{proposition}

\vspace{0.1cm} The above two propositions provide a complete
algorithm, in four steps, to detect and characterize, in any given
family of divergence-free perfect energy tensors ${\bf T} = \{{\rm
T} \equiv [u^{\alpha}(x^{\beta}), \rho(x^{\beta}), p(x^{\beta})]
\}$, those that represent an ideal gas evolving in local thermal
equilibrium:

\begin{description}
\item[step 1:] Calculate the coordinate dependence of the
space-time functions $\  p/\rho =\pi(x^{\beta})$ and $\
u(p)/u(\rho) = \chi(x^{\beta})$
\item[step 2:] Determine the ideal gas subset of ${\bf T}$
by imposing the ideal gas  hydrodynamic condition
(\ref{dchi-wedge-dpi}): $\ \ d \chi \wedge d \pi = 0$.
\item[step 3:] Obtain, in this subset, the explicit expression of the
indicatrix function: $\ \ \chi=\chi(\pi)$.
\item[step 4:] Calculate, from $\chi(\pi)$, the {\em generating
functions} $e(\pi)$ and $f(\pi)$ given in (\ref{e-pi}) and
(\ref{f-pi}), and obtain from them the thermodynamic variables by
using (\ref{T}) and (\ref{r-s}).
\end{description}

\section{\large THERMODYNAMIC STEPHANI UNIVERSES}

The conformally flat perfect energy tensor solutions to Einstein
equations with nonzero expansion were first considered by Stephani
\cite{st} and are usually called {\em Stephani universes}. They
were rediscovered by Barnes \cite{ba}, who obtained them as the
conformally flat class of irrotational and shear-free perfect
fluid space-times with nonzero expansion.

In order to generalize the cosmological principle, Bona and Coll
\cite{bc} looked for space-times admitting an iso-invariant
synchronization, and without any hypothesis on the energy tensor,
they likewise found the Stephani universes.

An iso-invariant synchronization amounts the existence of an
irrotational and shear-free observer $u$ with non zero expansion
such that the induced metrics on the orthogonal hypersurfaces are
of constant curvature. Then, there exists an adapted coordinate
system, $u= (1/\alpha)\partial_t$, such that the line element
takes the form:
\begin{equation}
\label{dsdos}
d{\rm s}^2 = -\alpha^2 dt^2 + \Omega^2 \delta_{ij} dx^i dx^j
\end{equation}
where
\begin{equation}
\displaystyle{\alpha \equiv R \partial_R \ln \Omega \qquad ,
\qquad \Omega \equiv \frac{R(t)}{1+ 2 \vec{b}(t) \cdot \vec{{\rm
r}} + \frac{1}{4} K(t) {\rm r}^2}}
\label{eq:metric}
\end{equation}
$R(t)$, $\vec{b}(t)$ and $K(t)$ being five arbitrary functions of
time.

The metric (\ref{dsdos}) is a solution to the perfect energy
tensor of the form (\ref{energytensor}), with energy density and
pressure given by
\begin{equation}
\rho = \frac{3}{R^2}(\dot{R}^2 + K - 4b^2) \qquad , \qquad p= -
\rho - {R \over 3} {\partial_R\rho \over \alpha} \label{eq:dp}
\end{equation}
Moreover, the expansion of $u$ and the curvature of the spatial
metric are homogeneous and  respectively given by
\begin{equation}  \label{curvatura}
\theta(t) = {3 \dot{R} \over R} \not= 0 \; ,
\qquad    \qquad \kappa(t)= {1 \over R^2} (K-4b^2) \end{equation}

The conditions under which Stephani universes describe the
evolution of a fluid in local thermal equilibrium, were directly
obtained by Bona and Coll in \cite{bc}, and later considered by
Krasi\'nski et al. \cite{ks}. The results that we need from
\cite{bc} are here summarized in the following lemmas.

\begin{lemma} The necessary and sufficient condition for a Stephani
universe to represent the evolution of a fluid in local thermal
equilibrium is that it admit a three-dimensional isometry group.
\end{lemma}

A particular class of evolutions in local thermal equilibrium is
that of {\em barotropic evolutions}; in our case, they are easily
determined:

\begin{lemma} A Stephani universe represents the barotropic
evolution of a fluid iff it admits a six-dimensional isometry
group (Friedmann-Robertson-Walker space-time).
\end{lemma}

The following lemma states, in thermodynamical terms, that
Stephani universes do not admit isometry groups of dimension 4 or
5.

\begin{lemma}The dimension of the maximal isometry group of a
Stephani universe that represents the non barotropic evolution of
a fluid is three.
\end{lemma}

This paper deals with Stephani universes describing non barotropic
evolutions of fluids in local thermal equilibrium. In what
follows, they will be called for short {\em thermodynamic Stephani
universes}.

A canonical form for them has also been obtained in \cite{bc}.
From it, and matching the two space-like coordinates ${\rm r}$ and
$z$ in a sole variable, say $w$, a straightforward calculation
leads to:
 \begin{proposition}
The metric of the thermodynamic Stephani universes may be written
\begin{equation}  \label{the-ste-uni}
d{\rm s}^2 = -\alpha^2 dt^2 + \Omega^2 \delta_{ij} dx^i dx^j
\end{equation}
where
\begin{equation} \label{eq:termetric}
\begin{array}{c}
\displaystyle \alpha \equiv R \partial_R \ln L \qquad , \qquad
\Omega \equiv \frac{w}{2z} L\\[2mm]
\displaystyle L \equiv \frac{R(t)}{1+ b(t) w} \qquad ,   \qquad w
\equiv \frac{2z}{1 + {\epsilon \over 4}{\rm r}^2}
\end{array}
\end{equation}
$R(t)$ and $b(t)$ being two arbitrary functions of time.

Its symmetry group is spherical, plane or pseudospherical
depending on $\varepsilon$ to be $1$, $0$ or $-1$ and the
Friedmann-Robertson-Walker limit occurs when $b=constant$.

Furthermore, the energy density and pressure are given by
\begin{equation}
\rho = \frac{3}{R^2} (\dot{R}^2 + \varepsilon - 4 b^2)  \qquad ,
\qquad p= - \rho - {R \over 3} {\partial_R\rho \over \alpha}
\label{tdp}
\end{equation}
\end{proposition}

\section{\large IDEAL GAS STEPHANI UNIVERSES}

Among the thermodynamic Stephani universes, what are those
corresponding to ideal gases?

In order to determine them, we shall study the restrictions that
our above ideal gas condition (\ref{dchi-wedge-dpi}) imposes on
the functions $R(t)$ and $b(t)$ that characterize the general
thermodynamic Stephani universes
(\ref{the-ste-uni})\,(\ref{eq:termetric}), following step by step
the algorithm presented at the end of section 2.

\subsection*{step 1: {\em Coordinate dependence of $\pi$ and $\chi$}}

From the second equation in (\ref{tdp}),
 a direct calculation leads to:
\begin{equation}
\label{eq:pi0}
  \frac{p}{\rho} =\pi(R, w)= \frac{a}{\alpha} -1
\end{equation}
where $\alpha$ is given in (\ref{eq:termetric}) and $a$ is defined by
\begin{equation}
  a  =  a(R) \equiv  - \frac{R\, \partial_R\rho}{3 \rho}  \ \ \ ,
\label{eq:a}
\end{equation}

Now we look for the indicatrix function $\chi$. From (\ref{eq:pi0})
and (\ref{eq:a}), we have
\begin{equation}
\label{eq:chi0}
 \chi \equiv  \frac{u(p)}{u(\rho)} = \frac{\partial_R p}{\partial_R
 \rho} = \frac{1}{\partial_R \rho} \partial_R (\rho \pi) = \pi -
 \frac{R}{3a} \partial_R \left(\frac{a}{\alpha}\right)
\end{equation}
but, from the definition of $\alpha$ in (\ref{eq:termetric}) it
follows
\begin{equation}
  \alpha  =  \alpha(R,w)  \equiv  \frac{1+(b-Rb')w}{1+bw},
\label{eq:alpha}
\end{equation}
where $x'$ stands for $dx/dR$ for any $x = x(R).$ Then, from
(\ref{eq:pi0}) and (\ref{eq:alpha}) it results:
$$
\partial_R \alpha = \left(\frac{a}{\pi+1} - 1 \right)
\left(\frac{b''}{b'} + \frac{a}{R(\pi+1)}\right)
$$
and substituting in (\ref{eq:chi0}), we obtain:
\begin{lemma}
For the thermodynamic Stephani universes, the thermodynamic
variable $\pi \equiv p/\rho$ and the indicatrix function $\chi
\equiv u(p)/u(\rho)$ take the expressions:
\begin{eqnarray}
  \pi & = & \pi(R, w) = \frac{a(1+bw)}{1+(b-Rb')w}-1 \qquad ,
\label{eq:pi}  \\
  \chi & = & \chi(\pi,R) \equiv  \pi + \frac{1}{3} -
  \frac{1}{3}(\pi+1)[\frac{a'R}{a^2}+\frac{1}{a}+
  (\pi+1-a)\frac{Rb''}{a^2b'}] \quad \quad \label{eq:chi}
\end{eqnarray}
where $a$ is given by {\rm (\ref{eq:a})} and the prime indicates
derivative with respect to the variable $R$.
\end{lemma}

\subsection*{step 2: {\em Ideal gas hydrodynamic condition}: $\; d \chi
\wedge d \pi = 0$}

Expression (\ref{eq:chi}) shows that, in agreement with the local
thermal equilibrium condition, the indicatrix function $\chi$
depends on the only energetic variables $\rho$ and $p$, $\chi =
\chi(\rho,p)$. Indeed, (\ref{eq:chi}) can be written:
\begin{equation}  \label{eq:chi2}
\chi = \chi(\pi,R) \equiv  \pi + \frac{1}{3} +
  \frac{1}{3}(\pi+1)[(\pi+1)A_1(R) + A_2(R)] \quad \quad
\end{equation}
where $\pi= p/\rho$ and where
\begin{equation}
\label{eq:A1A2}
  A_1(R) \equiv -\frac{Rb''}{a^2b'} \, , \qquad A_2(R) \equiv
  \frac{Rb''}{ab'} - \frac{a'R}{a^2} - \frac{1}{a}
\end{equation}
Thus, $\rho$ being an effective function of $R$, the functions
$A_1$ and $A_2$ can be considered as depending on $\rho$. Every
choice of these two functions determines different indicatrix
functions that correspond to different media.  How we must take
$A_1$ and $A_2$ in the ideal gas case?

The thermodynamic Stephani universes describing non barotropic
evolutions of fluids in local thermal equilibrium are such that
the function $\pi(R,w)$ in (\ref{eq:pi}) satisfies $\partial_w \pi
\not=0$. Consequently condition $\; d \chi \wedge d \pi = 0$ is
equivalent to $\partial_R \chi(\pi,R)=0$. From (\ref{eq:chi2})
this equation may be written:
\begin{equation}
\label{eq:xipi1}
  A_1'(R)(\pi+1) + A_2'(R) = 0
\end{equation}
with $A_1(R)$ and $A_2(R)$ given by (\ref{eq:A1A2}).

But, $\pi$ and $R$ being independent variables, from
(\ref{eq:xipi1}) it follows that the functions in (\ref{eq:A1A2})
are constant, $A_1(R)= c_{\scriptscriptstyle 1}$, $A_2(R)=
c_{\scriptscriptstyle 2}$. From now on, $c_{\scriptscriptstyle 1}$
and $c_{\scriptscriptstyle 2}$ will be called the {\em principal
constants.}

Moreover we can eliminate the function $b(R)$ in the second of
equations (\ref{eq:A1A2}) by using its values from the first one.
Thus, in a first step, we obtain two equations for the two
functions $b(R),$ and $\rho(R)$, this last one being related to
$a(R)$ by (\ref{eq:a}). In a second step, we put every solution
$b(R),$ $\rho(R)$ for these equations in the expression
(\ref{tdp}) of the energy density $\rho$ in order to obtain an
equation for the expansion factor $R(t)$, which determines then
the corresponding Stephani universe.  More precisely, we can
state:
\begin{theorem} \label{theo:igsu}
A thermodynamic Stephani universe {\rm (\ref{the-ste-uni})}\,{\rm
(\ref{eq:termetric})} represents  an ideal gas if, and only if,
the metric function $b(R)$ and the energy density $\rho(R)$
satisfy the two equations:
\begin{equation} \label{eq:eq12}
R a' = -a(c_{\scriptscriptstyle 1}a^2+c_{\scriptscriptstyle 2}a+1)
\, , \qquad \quad b'' R = - c_{\scriptscriptstyle 1} a^2 b'
\end{equation}
where a(R) is given in {\rm (\ref{eq:a})} and the principal
constants $c_{\scriptscriptstyle 1}$ and $c_{\scriptscriptstyle
2}$ are
arbitrary. \\[1mm]
Then, in terms of $b(R)$ and $\rho(R),$ the expansion factor
$R(t)$ satisfies the generalized Friedmann equation:
\begin{equation} \label{eq:Friedmann}
\rho(R) = \frac{3}{R^2}[ \dot{R}^2 + \varepsilon - 4 b^2(R)]
\end{equation}
\end{theorem}
\vspace{1mm}

A first glance at equations (\ref{eq:eq12})\,(\ref{eq:Friedmann})
allows us to know that the ideal gas Stephani universes are,
generically, a family of solutions depending on seven parameters,
namely: the two principal constants $c_{\scriptscriptstyle 1}$ and
$c_{\scriptscriptstyle 2}$ (we will see below that they determine
five classes with different thermodynamical properties), the
initial values $a_{\scriptscriptstyle 0}$,
$\rho_{\scriptscriptstyle 0}$ and $R_{\scriptscriptstyle 0}$ for
the functions $a(R)$ $\rho(R)$ and $R(t)$ and, finally, two
integration constants $b_{\scriptscriptstyle 1}$ and
$b_{\scriptscriptstyle 2}$ giving the general solution $b(R)$ to
the second equation in (\ref{eq:eq12}). In next section we study
in detail equations (\ref{eq:eq12})\,(\ref{eq:Friedmann}) and we
will analyze the properties of the models that we obtain depending
on these seven parameters.

\subsection*{step 3: {\em Indicatrix function}: $\chi=\chi(\pi)$}

Now we look for the explicit expression of the indicatrix
function. Assuming equations (\ref{eq:eq12}), the indicatrix
(\ref{eq:chi}) becomes a second degree polynomial in $\pi$ whose
coefficients depend on the principal constants
$c_{\scriptscriptstyle 1}$ and $c_{\scriptscriptstyle 2}$. More
precisely, we have:
\begin{proposition}  \label{pro:indicatrix}
The indicatrix function $\chi(\pi)$ of the ideal gas Stephani
universes is of the form:
\begin{equation} \chi(\pi) =
  \beta \pi^2 + \gamma \pi + \delta \, ,
\label{eq:chipi}
\end{equation}
where $\beta$, $\gamma$ and $\delta$ are constants related to
$c_{\scriptscriptstyle 1}$ and $c_{\scriptscriptstyle 2}$ by
\begin{equation}
\label{constants}
 \beta = \frac{c_{\scriptscriptstyle 1}}{3} \, , \qquad \quad
 \gamma= 1 + \frac{1}{3}(c_{\scriptscriptstyle
 2}+2c_{\scriptscriptstyle 1}) \, ,   \qquad \quad
 \delta = \frac13 (c_{\scriptscriptstyle 1}+
 c_{\scriptscriptstyle 2}+1) \, .
\end{equation}
\end{proposition}
\subsection*{step 4: {\em Thermodynamic variables}}

Finally, the expression (\ref{eq:chipi}) of the indicatrix
function $\chi(\pi)$ determines, according to proposition 2, the
other thermodynamic variables, and so, the thermodynamical
properties of the fluid. Expressions (\ref{T}) and (\ref{r-s})
show that the thermodynamic variables are determined by the
generating functions $e(\pi)$ and $f(\pi)$ given by (\ref{e-pi})
and (\ref{f-pi}) respectively. Consequently, the thermodynamic
scheme depends on the {\em reduced indicatrix} $\bar{\chi}$:
\begin{equation}
 \bar{\chi}(\pi) \equiv \chi(\pi)-\pi =
 \beta \pi^2 + \bar{\gamma} \pi + \delta \, ,
\label{eq:chibarpi}
\end{equation}
with $\bar{\gamma} \equiv \gamma -1$, and $\beta$, $\gamma$ and
$\delta$ given by (\ref{constants}). Then, depending on the
degree and the roots of $\bar{\chi}(\pi)$ we find five different
classes of ideal gas Stephani universes and, for every class,
the integrals (\ref{e-pi}) and (\ref{f-pi}) admit simple
analytic expressions. We summarize the results in the following:
\begin{proposition} \label{pro:tsch}
From a thermodynamical point of view and according to the values
of the principal constants $c_{\scriptscriptstyle 1}$ and
$c_{\scriptscriptstyle 2}$, the ideal gas Stephani universes may
belong to five classes. For every one of them, the reduced
indicatrix $\bar{\chi}(\pi)$ and the generating functions $e(\pi)$
and $f(\pi)$ are given by
\begin{description}
\item {\sc Class 1 ($c_{\scriptscriptstyle 1} =
c_{\scriptscriptstyle 2} = 0$):}
  $\quad  \bar{\chi}(\pi) = \frac{1}{3}$
\begin{equation}
\label{sch1}
  f(\pi) = f_0 \exp\{3\pi\} \, , \qquad
  e(\pi) = \frac{e_0}{(\pi+1)^3} \exp\{3\pi\}
\end{equation}

\item {\sc Class 2 ($c_{\scriptscriptstyle 1} = 0, \
  c_{\scriptscriptstyle 2} \not= 0$):}
  $\quad  \bar{\chi}(\pi) = \frac13[(c_{\scriptscriptstyle 2} (\pi + 1)
  + 1]$
\begin{equation}
\label{sch2}
\begin{array}{l}
  \displaystyle f(\pi) = f_0 [(c_{\scriptscriptstyle 2} (\pi + 1) +
  1]^{\frac3{\scriptscriptstyle 2}} \\[4mm]
  \displaystyle e(\pi) = \frac{e_0}{(\pi+1)^3}[(c_{\scriptscriptstyle
  2} (\pi + 1) + 1]^{3(1+\frac1{{\scriptstyle c}_{\scriptscriptstyle
  2}})}
\end{array}
\end{equation}

\item {\sc Class 3 ($\Delta \equiv c_{\scriptscriptstyle 2}^2 -
4c_{\scriptscriptstyle 1} = 0, \ c_{\scriptscriptstyle 1}
\not=0$):}
  $  \quad \bar{\chi}(\pi) = \frac{{\displaystyle
  c}_{\scriptscriptstyle 2}}{12}[(c_{\scriptscriptstyle 2} (\pi + 1) +
  2]^2$
\begin{equation}
\label{sch3}
\begin{array}{l}
  \displaystyle f(\pi) = f_0
  \exp\left\{\frac{-12}{c_{\scriptscriptstyle 2}
  [c_{\scriptscriptstyle 2} (\pi + 1) + 2]}\right\}
\\[4mm]
  \displaystyle e(\pi) = e_0 \frac{[c_{\scriptscriptstyle 2} (\pi + 1)
+ 2]^3}{(\pi+1)^3}
  \exp\left\{\frac{-6(c_{\scriptscriptstyle
  2}+2)}{c_{\scriptscriptstyle 2}[c_{\scriptscriptstyle 2} (\pi + 1) +
2]}\right\}
\end{array}
\end{equation}

\item {\sc Class 4 ($\Delta \equiv c_{\scriptscriptstyle 2}^2 -
4c_{\scriptscriptstyle 1} > 0, \ c_{\scriptscriptstyle 1}
\not=0$):}
  $\quad \bar{\chi}(\pi) = \frac{c_{\scriptscriptstyle 1}}{3}(\pi -
  \pi_{+})(\pi - \pi_{-})$
\begin{equation}
\label{sch4}
\begin{array}{l}
  \displaystyle f(\pi) = f_0 \left(\frac{\pi - \pi_{+}}{\pi -
  \pi_{-}}\right)^{\frac{3}{\sqrt{\Delta}}} \, , \quad \pi_{\pm}
  \equiv -1 + \frac{1}{2c_{\scriptscriptstyle
  1}}(-c_{\scriptscriptstyle 2} \pm \sqrt{ \Delta})
\\[4mm]
  \displaystyle e(\pi) = e_0 \frac{(\pi - \pi_{+})^{\lambda_{+}}(\pi
  - \pi_{-})^{\lambda_{-}}}{(\pi+1)^3} \, , \quad \lambda_{\pm}
  \equiv \frac32 \left[1 \pm \frac{c_{\scriptscriptstyle
  2}+2}{\sqrt{\Delta}}\right]
\end{array}
\end{equation}

\item {\sc Class 5 ($\Delta \equiv c_{\scriptscriptstyle 2}^2 -
4c_{\scriptscriptstyle 1} < 0$):}
  $\ \  \bar{\chi}(\pi) = \frac{1}{3}[c_{\scriptscriptstyle 1} \pi^2
  +(2c_{\scriptscriptstyle 1}+c_{\scriptscriptstyle 2}) \pi
  +c_{\scriptscriptstyle 1}+c_{\scriptscriptstyle 2}+1]$
\begin{equation}
\label{sch5}
\begin{array}{l}
  \displaystyle f(\pi) = f_0 \exp\left\{\frac{2}{3 \sqrt{-\Delta}}
  \arg\tan \frac{2c_{\scriptscriptstyle 1}(\pi+1) +
  c_{\scriptscriptstyle 2}}{\sqrt{-\Delta}}\right\}
\\[4mm]
  \displaystyle e(\pi) = e_0
  \frac{[\bar{\chi}(\pi)]^{3/2}}{(\pi+1)^3}
  \exp\left\{\frac{c_{\scriptscriptstyle 2}+2}{3
  \sqrt{-\Delta}} \arg\tan \frac{2c_{\scriptscriptstyle 1}(\pi+1) +
  c_{\scriptscriptstyle 2}}{\sqrt{-\Delta}}\right\}
\end{array}
\end{equation}
\end{description}
\end{proposition}
The complete space-time ideal models, which will be obtained in
next section by integrating equations (\ref{eq:eq12}), depend on
seven parameters. The above proposition shows that, among these
seven parameters, the principal constants $c_1$ and $c_2$ are
the two thermodynamical ones.

\section{\large IDEAL GAS STEPHANI MODELS}

The above section has obtained and explored the indicatrix
function $\chi$ for ideal gases in Stephani universes, and
constructed the generating functions $f(\pi)$ and $e(\pi)$ that
directly give the thermodynamic variables. Here we complete the
study by the exploration of the conditions (\ref{eq:eq12}) for the
existence of the indicatrix, which involve the metric functions.

The study of these conditions (\ref{eq:eq12}) leads to different
models that depend, generically, on seven parameters as pointed
out above. The principal constants $c_{\scriptscriptstyle 1}$ and
$c_{\scriptscriptstyle 2}$ distinguish the five classes of ideal
gas Stephani universes already obtained in proposition
\ref{pro:tsch}. But the analysis of the equations also
distinguishes two families of models defined by the fact that the
function $a(R)$ given in (\ref{eq:a}) may be constant or not. Now
we consider these two cases separately.

\subsection{Singular models: $a'(R) = 0$}

If $a(R)=a_{\scriptscriptstyle 0} \not= 0$, the first equation in
(\ref{eq:eq12}) implies that the parameter $a_{\scriptscriptstyle
0}$ of the model depends on the thermodynamic parameters, the
principal constants $c_{\scriptscriptstyle 1}$ and
$c_{\scriptscriptstyle 2}$. In these cases, we can easily
integrate (\ref{eq:a}) with $\;a(R) = a_{\scriptscriptstyle 0}$ to
obtain $\rho(R)$. On the other hand, the second equation in
(\ref{eq:eq12}) leads to $b' = C R^{-c_{\scriptscriptstyle
1}a_{\scriptscriptstyle 0}^2}$, $C$ being a constant. Then we
obtain:

\begin{theorem} \label{theo:models-a0}
In the singular ideal gas Stephani models, $a(R) =
a_{\scriptscriptstyle 0},$\\[1mm]
\indent i) the energy density is given by:
\begin{equation}
\label{ro-R-1} \rho(R)= \rho_{\scriptscriptstyle 0}
\left(\frac{R_{\scriptscriptstyle
0}}{R}\right)^{3a_{\scriptscriptstyle 0}}
\end{equation}
where $\rho_{\scriptscriptstyle 0}$ and $R_{\scriptscriptstyle
0}$ are constants, and \\[1mm]
\indent ii) the metric function $b(R)$ is given by one of the
following expressions
\begin{equation}
b(R) = b_{\scriptscriptstyle 1} + b_{\scriptscriptstyle 2}
\left(\frac{R}{R_{\scriptscriptstyle 0}}\right)^{\mu} \qquad if
\qquad
 \mu \not=0
\end{equation}
\begin{equation}
b(R) = b_{\scriptscriptstyle 1} + b_{\scriptscriptstyle 2} \ln
\left(\frac{R}{R_{\scriptscriptstyle 0}}\right) \qquad if \qquad
\mu =0 \quad .
\end{equation}
where $\mu \equiv 1-c_{\scriptscriptstyle 1} a_{\scriptscriptstyle
0}^2$, and $b_{\scriptscriptstyle 1}$, $b_{\scriptscriptstyle
2}\not=0$ are arbitrary constants.
\end{theorem}

As a consequence of (\ref{eq:eq12}), the relationship between the
constants $a_{\scriptscriptstyle 0},$ $\mu$ and
$c_{\scriptscriptstyle 1},$ $c_{\scriptscriptstyle 2}$ is given
by:
\begin{equation}
\label{ABamu} c_{\scriptscriptstyle 1} =
\frac{1}{a_{\scriptscriptstyle 0}^2} (1-\mu) \qquad , \qquad
c_{\scriptscriptstyle 2} = \frac{1}{a_{\scriptscriptstyle 0}}
(\mu-2)
\end{equation}
These expressions show that both, $a_{\scriptscriptstyle 0}$ and
$\mu$, are thermodynamical parameters. Moreover, one can easily
see how the above cases are related to the five classes of ideal
gas Stephani universes established in proposition \ref{pro:tsch}.

When $\mu =0$ one has $\Delta \equiv c_{\scriptscriptstyle 2}^2
-4c_{\scriptscriptstyle 1} = 0$, and these cases belong to class
3. When $\mu=1$ one has $c_{\scriptscriptstyle 1}=0$,
$c_{\scriptscriptstyle 2}\not=0$, and they belong to class 2.
All the other singular cases belong to class 4, because $\Delta
= (\mu/a_{\scriptscriptstyle 0})^2 > 0$. Thus, if we take into
account proposition \ref{pro:tsch} and (\ref{ABamu}) we can
state:

\begin{proposition}
\label{pro:sch-a0-models} The singular ideal gas models, $a(R) =
a_{\scriptscriptstyle 0},$ given in theorem {\rm
\ref{theo:models-a0}}, belong to the following classes:
\begin{description}
\item \hspace{3mm} if $\, \mu=1$, to class 2,
with $c_{\scriptscriptstyle 2} = - \frac{1}{a_{\scriptscriptstyle
0}}$.

\item \hspace{3mm} if $\, \mu=0$, to class 3, with
$c_{\scriptscriptstyle 2} = - \frac{2}{a_{\scriptscriptstyle 0}}$.

\item \hspace{3mm} if $\, 0 \not=\mu \not= 1$, to class 4, with
$c_{\scriptscriptstyle 1}$ and $c_{\scriptscriptstyle 2}$ given by
\rm{(\ref{ABamu})}.

\end{description}

\end{proposition}

\subsection{Regular models: $a'(R) \not= 0$}

Equations (\ref{eq:eq12}) are equivalent to equations
(\ref{eq:A1A2}) with $A_1=c_{\scriptscriptstyle 1}$ and $A_2 =
c_{\scriptscriptstyle 2}$; differentiating the first of these
equations and considering the second one, one finds:
\begin{equation}  \label{eq:k-a1}
\frac{a''}{a'} - \frac{2a'R}{a} -\frac{b''}{b'} = 0
\end{equation}
From this equation one can obtain $b$ as a function of $a$. But
this relation and the first equation in (\ref{eq:eq12}) imply the
second one. Thus, we have:

\begin{proposition}
The ideal gas Stephani universes with $a'(R) \not= 0$ are those
for which
\begin{equation} \label{eq:k-a2}
R a' = -a(c_{\scriptscriptstyle 1}a^2+c_{\scriptscriptstyle 2}a+1)
\, , \qquad \quad b(R) = \frac{b_{\scriptscriptstyle 2}}{a(R)} +
b_{\scriptscriptstyle 1}
\end{equation}
where $c_{\scriptscriptstyle 1}$, $c_{\scriptscriptstyle 2}$,
$b_{\scriptscriptstyle 1}$ and $b_{\scriptscriptstyle 2}\not=0$
are arbitrary constants.
\end{proposition}

Consequently, for regular ideal gas models, one needs to solve the
first order differential equation (\ref{eq:k-a2}) for the function
$a(R)$. The solutions can be obtained in an implicit form
$h(a,R)=0$ depending on the degree and roots of the polynomial
$q(a) \equiv c_{\scriptscriptstyle 1}a^2+c_{\scriptscriptstyle
2}a+1$. But an implicit expression for $a(R)$ does not allow to
determine explicit expressions for $\rho(R)$ (solution to
(\ref{eq:a})) and $b(R)$, and these expressions are necessary in
order to establish the Friedmann equation (\ref{eq:Friedmann}) and
to obtain the metric tensor by solving it. We can overcame this
shortcoming by a change of variables. Indeed, $a$ been an
effective function of $R$, the equation (\ref{eq:a}) can be
written
\begin{equation}
\frac{1}{\rho}\frac{d \rho}{d a} = - \frac{3a}{R}\frac{d R}{d a}=
\frac{3}{c_{\scriptscriptstyle 1}a^2+c_{\scriptscriptstyle 2}a+1} \, ,
\end{equation}
and consequently, the energy density takes the expression
\begin{equation} \label{rho}
\rho(a) = \rho_{\scriptscriptstyle 0} Q^3(a)
\end{equation}
where
\begin{equation}
Q(a) \equiv \exp\left\{\int \frac{da}{q(a)}\right\} \, , \qquad
q(a) \equiv c_{\scriptscriptstyle 1}a^2+c_{\scriptscriptstyle 2}a+1
\end{equation}

On the other hand, the first equation in (\ref{eq:k-a2}) takes the
form $a q(a) \, d R/d a= R$, and its solutions depend on a simple
function of $q(a)$ and $Q(a)$. Finally, Friedmann equation
(\ref{eq:Friedmann}) may be considered as a differential equation
on $a(t)$ by changing:
\begin{equation}
\dot{R} = \frac{d R}{d a} \, \dot{a}= \frac{R}{a q(a)}\, \dot{a}
\end{equation}
All these considerations lead to the following:

\begin{theorem}  \label{theo:models}

The regular ideal gas Stephani models, $a'(R) \neq 0,$ are those
for which the metric functions $R$ and $b$ are of the form
\begin{equation}  \label{RK-a}
R(a) = \frac{R_{\scriptscriptstyle 0}}{a} \sqrt{|q(a)
Q^{c_{\scriptscriptstyle 2}}(a)|} \, , \quad \quad b(a) =
\frac{b_{\scriptscriptstyle 2}}{a} + b_{\scriptscriptstyle 1}
\end{equation}
where $Q(a)$ and $q(a)$ are given by
\begin{equation}
Q(a) \equiv \exp\left\{\int \frac{da}{q(a)}\right\} \, , \qquad
q(a) \equiv c_{\scriptscriptstyle 1}a^2+c_{\scriptscriptstyle 2}a+1
\end{equation}
and $a=a(t)$ is a solution to the generalized Friedmann equation
\begin{equation}
\label{eq:Friedmann2} \rho_{\scriptscriptstyle 0} Q^3(a) = {3
\dot{a}^2 \over a^2 q^2(a)} + {3 \over R^2(a)} [\varepsilon -
b^2(a)] \quad ,
\end{equation}
$c_{\scriptscriptstyle 1}$, $c_{\scriptscriptstyle 2}$,
$b_{\scriptscriptstyle 1}$ and $b_{\scriptscriptstyle 2}\not=0$
being arbitrary constants.
\end{theorem}

The analytic expression of the function $Q(a)$ depends on the
polynomial $q(a)$. Thus we can distinguish five different models
depending on the values of the constants $c_{\scriptscriptstyle
1}$ and $c_{\scriptscriptstyle 2}$. These cases coincide with the
five classes of ideal gas Stephani universes established in
proposition \ref{pro:tsch}. The integration of these five cases
leads to simple analytic expressions for the function $Q(a)$:

\begin{proposition}
\label{pro:r-models} According to the class to which the ideal gas
Stephani universe belongs, the functions $q(a)$ and $Q(a),$ that
determine by {\rm (\ref{RK-a})} the metric function $R$ when
$a'(R) \neq  0,$ are given by:
\begin{description}
\item {\sc Class 1 ($c_{\scriptscriptstyle 1} =
c_{\scriptscriptstyle 2} = 0$):} $\quad  q(a) = 1$
\begin{equation}
\label{model1}
Q(a) = \exp\{a\}
\end{equation}
\item {\sc Class 2 ($c_{\scriptscriptstyle 1} = 0, \
c_{\scriptscriptstyle 2} \not= 0$):} $\quad q(a) =
c_{\scriptscriptstyle 2}a+1$
\begin{equation}
\label{model2}
Q(a) = (c_{\scriptscriptstyle 2}a +1)^{1/c_{\scriptscriptstyle 2}}
\end{equation}
\item {\sc Class 3 ($\Delta \equiv c_{\scriptscriptstyle 2}^2 -
4c_{\scriptscriptstyle 1} = 0, \ c_{\scriptscriptstyle 1}
\not=0$):} $ \quad q(a) = \frac14(c_{\scriptscriptstyle
2}a+2)^2$
\begin{equation}
\label{model3} Q(a) = \exp\left\{\frac{4}{c_{\scriptscriptstyle 2}
(c_{\scriptscriptstyle 2}a+2)}\right\}
\end{equation}
\item {\sc Class 4 ($\Delta \equiv c_{\scriptscriptstyle 2}^2 -
4c_{\scriptscriptstyle 1} > 0, \ c_{\scriptscriptstyle 1}
\not=0$):} $\quad q(a) = c_{\scriptscriptstyle 1}(a - a_{+})(a -
a_{-})$
\begin{equation}
\label{model4} \displaystyle Q(a) = \left(\frac{a - a_{+}}{a -
a_{-}}\right)^{\frac{1}{\sqrt{\Delta}}} \, , \quad a_{\pm} \equiv
\frac{1}{2c_{\scriptscriptstyle 1}}(-c_{\scriptscriptstyle 2} \pm
\sqrt{ \Delta})]
\end{equation}
\item {\sc Class 5 ($\Delta \equiv c_{\scriptscriptstyle 2}^2 - 4c_{\scriptscriptstyle 1} < 0$):}
$\quad q(a) =c_{\scriptscriptstyle 1} a^2+c_{\scriptscriptstyle
2}a+1$
\begin{equation}
\label{model5} \displaystyle Q(a) =
\exp\left\{\frac{2}{\sqrt{-\Delta}} \arg\tan
\frac{3(2c_{\scriptscriptstyle 1}a + c_{\scriptscriptstyle
2})}{\sqrt{-\Delta}}\right\} \quad.
\end{equation}
\end{description}
\end{proposition}

\section{\large APPROXIMATE CLASSICAL IDEAL GAS}

Until now we have considered {\em generic} ideal gases
characterized by the equation of state $p = k r T$. In this
section we consider {\em classical} ideal gases which also satisfy
the energetic equation of state $\epsilon = c_v T$, and we analyze
when a Stephani universe represents a classical ideal gas in local
thermal equilibrium. This energetic equation of state restricts
the indicatrix function $\chi(\pi)$ of the ideal gases. More
precisely, we have shown elsewhere \cite{cf3}:

\begin{proposition} \label{pro:clas1}
The indicatrix function of a classical ideal gas takes the
expression:
\begin{equation} \label{clas1}
\chi (\pi) = \frac{\gamma_a \pi}{1+ \pi}
\end{equation}
where $\gamma_a = 1 + \frac{k}{c_v}$, is the adiabatic index, $1
< \gamma_a < 2$.\\[1mm]
If a perfect fluid has the indicatrix function {\rm
(\ref{clas1})}, then it represents a classical ideal gas with
specific energy:
\begin{equation}  \label{clas1b}
e(\pi) = \frac{\gamma_a-1}{\gamma_a-1-\pi}
\end{equation}
\end{proposition}

From this result and proposition \ref{pro:indicatrix} we can
state: {\em a thermodynamic Stephani universe never represents in
a exact way a classical ideal gas in local thermal equilibrium}.
Nevertheless we should take into account that, in a relativistic
framework, the classical ideal gas approximation works for low
temperatures. Thus, we can look for ideal gas schemes that satisfy
the energetic equation of state of a classical ideal gas at first
order in $T$. This fact can be analyzed by studying the
expressions (\ref{T}) and their first and second derivatives for
small $T$. Then, taking into account (\ref{e-pi}) and proposition
9, we obtain

\begin{proposition}
The necessary and sufficient condition for an ideal gas to have,
at first order in the temperature, the energetic equation of state
of a classical ideal gas, $\epsilon(T)= c_v  T + o(T^2)$, is that
the indicatrix function $\chi(\pi)$ and the specific energy
$e(\pi)$ satisfy:
\begin{equation} \label{chi0}
\chi(0) = 0 \, , \qquad  \chi'(0) = \gamma_a \, ; \qquad \quad
e(0) = 1
\end{equation}
\end{proposition}

Now we can study the compatibility between (\ref{chi0}) and the
Stephani ideal gas indicatrix function (\ref{eq:chipi}) at first
order in $\pi$, and we easily obtain $\delta = 0$ and $\gamma =
\gamma_a$. Moreover, this means that the adiabatic index
$\gamma$ fixes the parameters $c_{\scriptscriptstyle 1}$ and
$c_{\scriptscriptstyle 2}$ of the classical ideal gas models.
More precisely we have:

\begin{proposition}  \label{pro:clas2}
If an ideal gas Stephani universe represents a classical one at
first order in the temperature, then the indicatrix function can
be written:
\begin{equation}  \label{clas2}
\chi(\pi) = (\gamma-\frac23 ) \pi^2 + \gamma \pi \, , \qquad
1<\gamma<2
\end{equation}
and the principal constants are the following functions of
$\gamma$:
\begin{equation}  \label{clas2b}
c_{\scriptscriptstyle 1} = 3 \gamma - 2 > 1 \, , \qquad
c_{\scriptscriptstyle 2} = 1 - 3 \gamma  < -2
\end{equation}
\end{proposition}

This proposition implies that, if we impose on the thermodynamic
parameters $c_1$ and $c_2$ the conditions (\ref{clas2b}), the
ideal gas indicatrix function $\chi = \chi(\pi)$ satisfies
(\ref{chi0}). Then, only the models of class 4 in proposition
\ref{pro:tsch} are admitted in this case. Indeed, from
(\ref{clas2b}) we obtain:

\begin{equation}  \label{clas-delta}
c_{\scriptscriptstyle 1} \not= 0 \, , \qquad \Delta =
(c_{\scriptscriptstyle 1}-1)^2 = 9(\gamma-1)^2 >0
\end{equation}
which agree with the conditions of proposition \ref{pro:tsch} that
define the ideal gas Stephani universes of class 4. Moreover, if
we determine the parameters $\pi_{\pm}$ and $\lambda_{\pm}$ of the
scheme (\ref{sch4}) in terms of the adiabatic index $\gamma$ and
we fix the constant $e_0$ in order to have $e(0) = 1$, we can
state:

\begin{proposition}
For the Stephani universes, the ideal gas thermodynamic schemes
which approximate, at first order in the temperature, a classical
one with adiabatic index $\gamma$, are generated by the functions
\begin{equation}
f(\pi) = f_0 \left[\frac{\pi}{( \gamma -2/3) \pi +
(\gamma-1)}\right]^{\frac{1}{\gamma-1}} \! \! ,  \ \ e(\pi)
=\left[1+\frac{\pi}{3(\gamma-1) (\pi + 1)}\right]^{3}
\end{equation}
\end{proposition}

\subsection{Classical singular models}

As a consequence of proposition \ref{pro:sch-a0-models} the
singular models of class 4 satisfy $0 \not= \mu \not= 1$. Then,
the relations (\ref{ABamu}) and the restrictions (\ref{clas2b})
lead to either $a_{\scriptscriptstyle 0} = 1$ and $\mu =
3(1-\gamma)<0$, or $a_{\scriptscriptstyle 0}=1/(3\gamma-2)$ and
$\mu = 1-a_{\scriptscriptstyle 0}>0$. Then, if we take into
account theorem \ref{theo:models-a0}, we can state:

\begin{proposition}
There are two families of singular models that approximate, at
first order in the temperature, a classical ideal gas with
adiabatic index $\gamma$:
\begin{description}
\item
{\sc Models $a_{\scriptscriptstyle 0}=1$}:
\begin{equation} \label{ro-K-clas1}
\rho(R)= \rho_{\scriptscriptstyle 0}
\left(\frac{R_{\scriptscriptstyle 0}}{R}\right)^{3} \, , \qquad
b(R) = b_{\scriptscriptstyle 1} + b_{\scriptscriptstyle 2}
\left(\frac{R_{\scriptscriptstyle 0}}{R}\right)^{3(\gamma-1)}
\end{equation}

\item
{\sc Models $a_{\scriptscriptstyle 0}\not=1$}:
\begin{equation} \label{ro-K-clas2}
\rho(R)= \rho_{\scriptscriptstyle 0}
\left(\frac{R_{\scriptscriptstyle
0}}{R}\right)^{\frac{1}{\gamma-2/3}} \, , \qquad b(R) =
b_{\scriptscriptstyle 1} + b_{\scriptscriptstyle 2}
\left(\frac{R}{R_{\scriptscriptstyle
0}}\right)^{\frac{\gamma-1}{\gamma-2/3}}
\end{equation}
\end{description}
\end{proposition}

\subsection{Classical regular models}

Proposition \ref{pro:r-models} and relations (\ref{clas-delta})
imply that the models of class 4 are the sole regular models
compatible with the classical ideal gas approximation in question.
If we write the parameters of the model in terms of $\gamma$ by
using (\ref{clas2b}), we obtain the following result:

\begin{proposition}
For the regular models in theorem {\rm \ref{theo:models}} that
approximate, at first order in the temperature, a classical ideal
gas with adiabatic index $\gamma$, the functions $q(a)$, $Q(a)$,
and $R(a)$ take the expression:
\begin{equation}
\begin{array}{l}
\displaystyle q(a) = (a-1)[(3\gamma-2)a-1] \, , \qquad  Q(a) =
\left[\frac{a-1}{(3\gamma-2)a -1}\right]^\frac{1}{3(\gamma-1)} \,
,
\\[4mm]
\qquad \qquad \quad \displaystyle  R(a) =
\frac{R_{\scriptscriptstyle 0}}{a} \left[\frac{[(3\gamma-2)a
-1]^{3\gamma-2}}{a-1}\right]^\frac{1}{3(\gamma-1)}
\end{array}
\end{equation}
\end{proposition}

\section{\large ON THE PHYSICAL BEHAVIOR OF THE MODELS}

In this work we have presented Stephani models that describe
evolutions in local thermal equilibrium of a generic ideal gas
(section 5) or a classical ideal gas (section 6). The space-time
domain where these models have a good physical behavior are
defined by complementary physical requirements.

Firstly, we have the Pleba\'nski energy conditions that, for a
perfect fluid, state $\, - \rho < p \leq \rho \, $. In terms of
the hydrodynamic variable $\pi = p/\rho$ these conditions take the
form $\, - 1 < \pi \leq 1 \, $. Nevertheless, in the ideal gas
case it is reasonable to consider positive pressures $p$. Thus we
should impose on $\pi$:
\begin{equation}
0 < \pi \leq 1
\end{equation}

On the other hand, the relativistic compressibility conditions
\cite{lich} should be also required. Elsewhere \cite{cf} we have
shown that these conditions can be written in terms of
hydrodynamic variables by means of the indicatrix function
$\chi(\rho,p)$, and that for an ideal gas they become \cite{cf}:
\begin{equation}  \label{comcon}
\frac{\pi}{2\pi+1} < \chi(\pi) < 1 \, , \qquad \chi'(\pi) > -
\frac{2 \chi(\pi) (1- \chi(\pi))}{(\chi(\pi) - \pi) (\pi + 1)}
\end{equation}

These conditions imposed on the indicatrix function
(\ref{eq:chipi}) of the ideal gas Stephani universes imply the
restriction of the parameters $c_{\scriptscriptstyle 1}$ and
$c_{\scriptscriptstyle 2}$ which determine the thermodynamical
properties of the models. Moreover, given a pair of values
$(\tilde{c}_{\scriptscriptstyle 1}, \tilde{c}_{\scriptscriptstyle
2})$ physically compatible, the compressibility conditions
(\ref{comcon}) only hold for values of $\pi$ on a subset of the
interval $]0,1]$. This means that the corresponding thermodynamic
scheme is only well defined for these values of $\pi$. Thus, for
example, for a model of class 1 (that is, $c_{\scriptscriptstyle
1} = c_{\scriptscriptstyle 2} = 0$), conditions (\ref{comcon})
hold if $\pi \in \; ]0, \frac23[$.

For a classical ideal gas the compressibility conditions
(\ref{comcon}) hold for every $\pi \in \;]0,1[$ if we assume that
the adiabatic index $\gamma \in \; ]1,2[$. Nevertheless, for the
Stephani models studied in section 6 which approximate a classical
ideal gas at first order in the temperature, these conditions only
hold for $\pi \in \; ]0, \pi_m[$, where the maximum value of $\pi$
depends on the adiabatic index as $\pi_m = \frac{2}{\gamma +
\sqrt{\gamma^2 + 4(\gamma - 2/3)}}<1$. It is worth remembering
that this approximation is valid only for small values of $\pi$.
Thus for $\pi$ near of the maximum value $\pi_{m}$, these models
have a good physical behavior but they do not approximate a
classical ideal gas.

Once obtained the admissible values of the hydrodynamic variable
$\pi$ for the different classes, we can look for the space-time
domains where every model is physically reasonable. We should
analyze the expression (\ref{eq:pi}) that gives $\pi$ in terms of
the spatial coordinate $\omega$ and the metric function $R(t)$. It
is worth remarking that $\omega$, as defined in
(\ref{eq:termetric}), is a bounded coordinate in the case of
spherical symmetry ($\varepsilon = +1$) and, otherwise, it is not
bounded. On the other hand, depending on the behavior of $R(t)$ we
could have models only valid at early or at present times. But all
these analysis will be considered elsewhere.

Finally, we want to comment about the relationship between the
results by Sussman in \cite{su} and ours. His models approximate a
classical ideal gas in some specific cases that, in our notation,
correspond to the spherically symmetric ($\varepsilon =1$)
singular case (\ref{ro-K-clas1}) with $a_{\scriptscriptstyle 0} =
1$ and $b_{\scriptscriptstyle 1} = 0$. In spite of this
restriction, Sussman's paper presents a wide analysis on the
space-time domains and observational features concerning the cases
of a classical mono-atomic ideal gas ($\gamma = 5/3$) and a
matter-radiation mixture ($\gamma = 4/3$).

\section*{\large ACKNOWLEDGMENTS}
This work has been supported by the Spanish Ministerio de Ciencia
y Tecnolog\'{\i}a, project AYA2003-08739-C02-02 (partially funded
by FEDER funds).


\begin{thebibliography}{999}
\bibitem{thermoclosures} All the causal closures that have
been proposed in the literature for arbitrary fluids are (or may
be related to) thermodynamical closures, originated in relativity
from many different approaches and giving rise to very different
results. They essentially started with Eckart's third paper on
thermodynamics of irreversible processes, and their contrast with
Landau and Lifchitz's point of view produced a great number of new
propositions, more or less decanted over the works by Israel,
Steward and Marle, and subsequently over the relativistic version
of extended thermodynamics.
\bibitem{perfect} By this and the other reasons mentioned above,
in the rest of the paper to the appellations `perfect fluid
space-times' or `perfect fluid solutions' (to Einstein equations),
we prefer those of `perfect energy tensor space-times' or `perfect
energy tensor solutions'.
\end{thebibliography}
\end{document}